\documentstyle[11pt,newpasp]{article}

\begin{document}

\title{Analytical Properties of the R$^{1/m}$ Law}

\author{L. Ciotti\altaffilmark{1} and G. Bertin}
\affil{Scuola Normale Superiore, Piazza dei Cavalieri 7, I-56126 Pisa}


\altaffiltext{1}{Osservatorio Astronomico di Bologna, via Ranzani 1, 
                 I-40127 Bologna}



\begin{abstract}
In this paper we describe some analytical properties of the $R^{1/m}$
law proposed by Sersic (1968) to categorize the photometric profiles
of elliptical galaxies.  In particular, we present the full asymptotic
expansion for the dimensionless scale factor $b(m)$ that is introduced
when referring the profile to the standard effective radius.
Surprisingly, our asymptotic analysis turns out to be useful even for
values of $m$ as low as unity, thus providing a unified analytical
tool for observational and theoretical investigations based on the
$R^{1/m}$ law for the entire range of interesting photometric
profiles, from spiral to elliptical galaxies.  The systematic
asymptotic analysis provided here also allows us to clarify the value
and the limitations of the power law $R^{-2}$ often used in the past
as a natural representation of the surface brightness profiles of
elliptical galaxies.
\end{abstract}




\section{Introduction}

After its introduction as a generalization of the $R^{1/4}$ law (de
Vaucouleurs 1948), the $R^{1/m}$ law (Sersic 1968) has been widely
used in observational and theoretical investigations (see, e.g.,
Ciotti \& Bertin 1999, and references therein, hereafter CB).
According to such law, the surface brightness profile is given by
\begin{equation}
I(R)=I_0 e^{-b\eta^{1/m}},\quad m>0,
\end{equation}
where $\eta=R/R_{\rm e}$, and $b$ is a dimensionless constant such
that $R_{\rm e}$ is the effective (half--luminosity) radius.  The
projected luminosity inside $R$ is given by
\begin{equation}
L (R)=2\pi\int_0^RI(R')R'dR'=I_0 R_{\rm e}^2 
                             {2\pi m\over b^{2m}}\gamma (2m,b\eta ^{1/m}),
\end{equation}
where $\gamma(\alpha,x)=\int_0^xe^{-t}t^{\alpha-1}dt$ is the (left)
incomplete gamma function. The total luminosity is then given by
\begin{equation}
L=I_0 R_{\rm e}^2 {2\pi m\over b^{2m}}\Gamma (2m),
\end{equation}
where $\Gamma(\alpha)=\gamma(\alpha,\infty)$ is the complete gamma
function.  It follows that $b(m)$ is the solution of the following
equation:
\begin{equation}
\gamma (2m,b) ={\Gamma (2m)\over 2}.
\end{equation}

\section{Asymptotic expansion}

Unfortunately, eq. (4) cannot be solved in explicit, closed form, and
so it is usually solved numerically.  This is inconvenient for a
number of observational and theoretical applications.
At least three interpolation formulae for $b(m)$ have been given in
the literature, namely $b\simeq 1.9992m-0.3271$ by Capaccioli (1989),
$b\simeq 2m-0.324$ by Ciotti (1991), and $b\simeq 2m-1/3+0.009876/m$
by Prugniel \& Simien (1997).  These expressions provide an accurate
fit in the range $0.5\leq m\leq 10$; curiously, their leading term is
{\it linear} in $m$, with a slope very close to 2.  In CB 
it is proved that this behavior results from a general
property of the gamma function.  In particular, it turns out that the first
terms of the asymptotic expansion of $b(m)$ as implicitly given by
eq. (4) are
\begin{equation}
b(m)\sim    2m-{1\over 3}
              +{4\over 405 m}
              +{46\over 25515 m^2}
              +{\rm O}(m^{-3}).
\end{equation}
Equation (5) now clearly explains the value of the interpolation
formulae found earlier.  Note that $4/405=0.009876\ldots\/$.  Of
course, the asymptotic analysis provided in CB would allow us to give
explicitly any higher order term if so desired.  In CB it is shown
that this expansion, even when truncated to the first four terms as in
eq. (5), performs much better than the interpolation formulae cited
above, even for $m$ values as low as unity, with relative errors
smaller than $10^{-6}$.

The use of this simple formula is thus recommended both in theoretical
and observational investigations based on the Sersic law. As a simple
application, in CB the leading term of the asymptotic expansion of the
total luminosity, of the central potential, and of the surface
brightness profile associated with the R$^{1/m}$ law are obtained in
explicit, analytical form.  In particular, the connection of eq. (1)
with the simple power law $R^{-2}$, often used in the past to fit the
photometric profiles of elliptical galaxies, is brought out
explicitly.

\acknowledgments
This work was supported by MURST, contract CoFin98. 


%
%

%


\begin{references}

\reference Capaccioli, M. 1989, in The world of galaxies, H.G. Corwin \& L.
           Bottinelli, Springer Verlag: New York, 208
\reference Ciotti, L., 1991, \astap, 249, 99
\reference Ciotti, L., \& Bertin, G. 1999, submitted (CB)
\reference de Vaucouleurs, G. 1948, Ann.d'Ap., 11, 247
\reference Prugniel, P., \& Simien, F. 1997, \astap, 321, 111
\reference Sersic, J.L. 1968, Atlas de galaxias australes. Observatorio
           Astronomico, Cordoba
\end{references}
\end{document}